# Designing Transformational Games to Support Socio-ethical Reasoning about Generative AI

Jaemarie Solyst*, Ruth Karen Nakigozi*, Chloe Fong, R. Benjamin Shapiro
jaemarie@cs.washington.edu, rnakig@uw.edu, cf2024@cs.washington.edu, rbs@cs.washington.edu
University of Washington

**Abstract:** There is an increasing need for young people to become critically AI literate, understanding not only how AI works but also its limitations and ethical nuances. Yet, designing learning experiences that make such complex, serious topics engaging remains a challenge. This paper explores *transformational games* as a promising approach for supporting youth learning about generative AI (GenAI) and ethics. We designed and implemented two games, Diversity Duel and Secret Agent, that integrate GenAI tools with gameplay elements. This work investigates how the games' elements: (1) peer evaluation, (2) constraint-based creativity, and (3) social deduction supported socio-ethical reasoning about GenAI. Participants recognized and debated bias in GenAI outputs, connected these patterns to real-world inequities, and developed nuanced understandings of bias. Participants further came to see how prompt design shapes AI behavior. Our findings suggest that group-based games with these elements can support fostering critical AI literacy.

## Introduction and relevant background

Artificial intelligence (AI) has rapidly become part of the everyday lives of youth, shaping how they learn, connect with peers and pop culture, consume media, and imagine futures. As AI systems increasingly mediate how information, creativity, and social interactions unfold, youth need opportunities to develop AI literacy–not only in using AI and understanding its workings (e.g., algorithms, data) but also in critically evaluating its limitations and socio-ethical implications (e.g., Touretzky et al., 2019; Long & Magerko, 2020; Ng et al., 2021). While generative AI (GenAI) holds great promise, many are concerned about detrimental impacts on critical thinking and social interaction; in a recent survey, almost half of teens reported they use it for companionship and one in five reported for romantic companionship (Laird et al., 2025). *Critical* AI literacy, i.e., understanding the capabilities and limitations of AI, is more important than ever. Yet, designing engaging and conceptually rich learning experiences remains challenging especially for complex or serious topics, such as GenAI ethics.

We see an opportunity for *transformational games* to be engaging and powerful mediums for supporting AI education. Transformational games are games that are intentionally designed to facilitate certain motivations, mindsets, learning, and sensemaking practices through low-stakes play (Culyba, 2018), e.g., by facilitating curiosity and normalizing making mistakes (To et al., 2018), roleplaying and perspective-taking for empathy building (Hammer et al., 2024) or group conversation (e.g., Corredor, 2018). GenAI offers unique affordances for creative interaction and production of new content, which may make it especially felicitous for game-based learning (Rafner et al., 2023). Prior work has shown that transformational games can support learning about AI (e.g., Ali et al., 2023; Hyo-Jeong & Sung-Eun, 2024; Showkat et al., 2025; Lim et al., 2025), but has not investigated transformational games for GenAI literacy in particular. Thus, there is great potential for games to support ethical and critical understanding of GenAI. For scope, we focus on text-to-image GenAI.

In this work, we introduce two transformational games designed to help youth engage with GenAI ethics through playful, social, and creative interaction, and describe their use within a broader out-of-school summer program on AI literacy. Then we present an analysis of how the game elements of (a) *competition with peer evaluation*, (b) *constraint-based creativity*, and (c) *social deduction* together support socio-ethical reasoning about bias in GenAI via gameplay. We describe our reasoning for these elements in the next section.

We found that youth who played the two games we designed, Diversity Duel and Secret Agent, engaged in rich, socially-mediated sensemaking about AI bias and ethics. Within gameplay, participants recognized visible forms of bias in GenAI outputs, connected these biases to real-world inequities, and collaboratively negotiated when bias might be acceptable or harmful. The games' playful, structured, and group formats enabled youth to reason about bias not as a fixed or purely technical problem, but as a complex socio-ethical issue shaped by human judgment and language. As they practiced careful prompt engineering, participants developed a more technical understanding of how linguistic specificity and prompt design affect AI behavior, seeing how every word could reproduce or challenge bias. Together, these findings suggest that group-based games can foster critical GenAI literacy by helping youth connect the elements of text-to-image AI generation with broader questions of ethics and human agency.

## Game design and learning goals

In response to abundant calls for youth to become critically AI literate, and the rise in popularity of technologies made with large language models, we focus on GenAI. Prior work on AI literacy and ethics for youth primarily frames *bias* as solely unfavorable (e.g., Lee et al., 2021; Solyst et al., 2025a), but bias–producing outputs that mirror socially produced associations between attributes of entities in the world– is necessary for AI to function or else systems would produce outputs at random (e.g., Kaelbling et al., 1996). Although AI can reproduce problematic societal biases, it has been debated (even by youth) whether or not AI should reflect current societal biases or an ideal future (Solyst et al., 2023). To develop critical AI literacy, one must engage with these nuanced concerns, developing the ability to go beyond recognizing them, to determine when specific biases are or are not so problematic, given particular contexts and applications of AI systems. In the following sections, we describe the learning goals we sought to support, the game elements we used, and the two games we created.

We designed our games to support learning of four key ethical practices for GenAI: (1) recognize bias in GenAI models, (2) recognize that GenAI biases reflect real-world biases, (3) understand that bias may be necessary, and (4) understand that some, but not all, bias can be harmful. At a high level, these learning goals are key to critical AI literacy, particularly for evaluating GenAI tools. The first game, Diversity Duel, emphasizes Learning Goals 1 and 2, while the second game, Secret Agent, emphasizes Learning Goals 3 and 4, although both games may support all four goals to an extent. Each game featured a different subset of those goals to support flexible learning experiences (e.g., shorter, simpler) and increased ability to understand how elements of play impact learning. We wanted to foster different interactions to meet these learning goals.

Transformational game designers invent or select game elements that encourage students to make sense of scenarios and make choices within them in ways that align with what designers want players to learn. Our two games employ three main gameplay elements: (a) *competition with peer evaluation*, (b) *constraint-based creativity*, and (c) *social deduction*. All three elements have been documented to support learning—with competition increasing motivation and learning gains (Cagiltay et al., 2015), including with peer-assessment (Thampy et al., 2023); constraints supporting deeper thinking and creativity (Tomp & Baer, 2022; Acar et al., 2019; Byron et al., 2010); and social deduction (e.g., like in popular games like *Mafia*) enhancing engagement and discussion (Tilton, 2019), including with difficult or intimidating science topics (Teychené & Dietrich, 2025; Bullock & Bloodsworth, 2025; Conner & Baxter, 2022). Using a large language model-powered image generator (DeepAI's AI Image Generator) was part of both games due to its ability to quickly create one image. In developing the games, we conducted multiple playtests with our team internally, improving the game designs.

## Game 1: Diversity Duel

In Diversity Duel, learners participate in independent groups of four, playing in pairs within each group. The game consists of three rounds. In each round, one player draws a card displaying an occupation prompt: "intelligent scholars," "construction workers," "teachers," "tech employees." We constructed the cards to elicit biases that we knew would be embedded in the GenAI systems. Learners use these cards as reference points to generate an image category according to the prompt. As shown in Figure 1, each pair shares a computer with a GenAI tool to create images for the round. The game stimulates **constraint-based creativity** by instructing pairs to write a short prompt aligning with the occupation on their card, using at most six words to generate an image that appears as "diverse" as possible. We define diverse as "including or involving people from a range of different social and ethnic backgrounds, and of different genders and sexual orientations." They have 45 seconds to collaboratively compose the prompt and are allowed up to two attempts to generate an image with six words and choose together the most diverse image generated. Next, the game sparks competition with **peer evaluation** by directing all four players to review and vote on which of the two images is more diverse. The pair whose image receives the most votes wins that round. In successive rounds, the word limit for prompts decreases (five, then four words), requiring players to explore using fewer prompting words that are effective for producing diverse results. Overall, the game combines competition with peer evaluation and constraint-based creativity to encourage learners to critically assess GenAI outputs together. They consider the impacts of words and human language used in prompts to generate diverse outputs.

**Figure 1**
*Overview of Diversity Duel's gameplay*

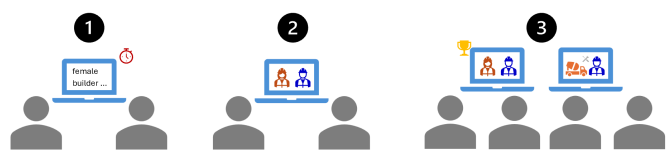

## Game 2: Secret Agent

Secret Agent is also a four-player game. Within each game, one player is randomly and covertly assigned the role of the "secret agent." The goal for the group is to collaboratively generate an inclusive image based on a prompt, like in Diversity Duel, while the secret agent's hidden objective is to subtly reduce the image's diversity by inserting words that may increase biases, as described in Figure 2. The game consists of two rounds, using the prompts, e.g., "construction workers" for the first and "tech employees" for the second. Each group of four shares a single computer with an image generator. For **constraint-based creativity**, players take turns and have thirty seconds each to add two words to a shared prompt. They pass the laptop along to each player until the complete prompt is formed and the image is generated. For **peer evaluation**, a separate group is asked to assess the image on two criteria: (1) whether it accurately represents the category (such as "construction workers"), and (2) whether it appears "diverse." This design is meant to encourage the secret agent to make their word choices subtle enough to avoid detection while still negatively influencing the image outcome. **Social deduction** occurs at the end of each round, where the group investigates the collaborative prompt, discusses, and identifies who they believe the secret agent is by reaching a majority vote. The secret agent earns a full win if the image is evaluated as not inclusive (by the other group) *and* they remain undetected, or a partial win if either (a) the image is not inclusive but they were caught, or (b) if they escaped detection despite the image being inclusive.

**Figure 2**
*Illustration of slides used in the broader summer study*

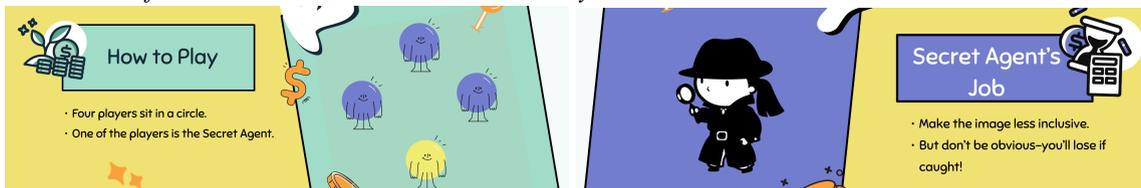

Examples of slides from the broader summer study illustrating the role of a Secret Agent.

## Design conjectures: How the gameplay elements should support learning

Because group sensemaking can enhance understanding of AI ethics (Salac et al., 2023; Landesman et al., 2024), we saw peer evaluation (in both games) and social deduction (in Secret Agent) as particularly fitting elements to support critical discourse about recognizing, critiquing, and understanding the potential role of bias in GenAI. Peer evaluation in Diversity Duel supports evaluation of negative bias in the image outputs, while peer evaluation in Secret Agent motivates critical discourse around how bias may be necessary to ensure that the output may be interpreted as the card drawn. In both games, the learners are tasked to create an output with constraint-based creativity: limiting the number of words that they can prompt with and restriction from using specific words directly, like "diverse" or "diversity." We see this as supporting the connection to prompt engineering, specifically how certain words can impact bias in image outputs. Through social deduction, the players in Secret Agent must think carefully about both injecting bias without being too obvious (the secret agent), and the other players must analyze and discuss the words in the prompt to figure out who the agent is.

## Methods

### Study context

We investigated our game design conjectures by using the games within a five day educational workshop held during summer break. The workshop focused on critical AI literacy, and Diversity Duel was played on Day 1 and Secret Agent on Day 2. We focus on these days for scope. We recruited study participants with the support of the community center that primarily centers youth of color; the workshop took place at a large research university. They suggested that we run the AI literacy workshop with girls, so all participants (N = 16) were teen girls aged 13-18 (average = 15.13). Eight identified as African American/Black, four identified as multiracial, three identified as Mexican, and one identified as White. We invited all youth who registered interest and met the age requirement (13 or older) to participate. Participants' computing education experiences varied: six learners reported that they had no prior experiences with computing education, while five learners had been involved with computing courses or robotics. The girls sat in pods of four, roughly by age, based on preferences.

On the first day, we conducted activities including icebreakers, so that the participants could get to know each other, and introductory activities about AI. For context, prior to the games, camp lessons included high-level definitions of AI and discussion about how it was a part of the learners' lives. We defined and briefly discussed the term diversity. On the second day, we explicitly solidified what they learned in Diversity Duel–we covered the concept of a stereotype prior to playing the Secret Agent game, in conjunction with the topic of algorithmic bias and how such biases appear in text-to-image GenAI outputs.

For both games, we administered pre- and post-questionnaires, designed to measure learning gains related to the learning goals. For the Diversity Duel, learners responded to three questions pertaining to slides showing sets of GenAI images, including twenty of "doctor" and "nurse" (ten each) with gender bias in the pre-questionnaire and then "executive" and "executive assistant" (ten each) in the post-questionnaire. The questions were: (1) "Do you think these are good images?" (Yes, No, or Unsure), (2) "Why is that?" (open-ended), and (3) "Why do you think the images were generated this way?" (open-ended). For the Secret Agent game, learners answered: (1) "Bias in AI is not harmful" (Strongly Disagree, Disagree, Neutral, Agree, Strongly Agree, or Unsure), and (2) "Why do you think that?" (open-ended). Participants completed the pre-questionnaire before gameplay and the identical post-questionnaire afterward. They also engaged in prompted group discussions after the post-questionnaires. For Diversity Duel, we asked learners to discuss: *Did this game change or not change the way you think about how AI creates images? Why (not)?* For Secret Agent, they considered the ethical dimensions of bias: *Can bias be helpful? Is it always harmful? Why (not)?*

### Data collection and analysis

We audio recorded group discussions and captured screen recordings of laptops during the games. We took pictures of any handwritten artifacts. Researcher notetakers also closely documented conversations during the activities. There was one trained research notetaker per learner pod. In total, we collected 253 minutes of recordings for Diversity Duel and 107 minutes for Secret Agent (the latter being shorter due to use of a single laptop shared per each group). We conducted consensus-based inductive thematic analysis (Braun & Clarke, 2006; Hammer & Berland, 2014) of all written worksheet responses, noted dialogue from learners, and researcher observation notes. Key segments from the audio and screen recordings were selectively reviewed based on emphasized moments in the notes. Two authors put together initial codes for each game, organized around the types of learning gains observed based on the pre and post questionnaires (e.g., better understanding, worse understanding, and no change). These codes were then grouped into subthemes and subsequently into broader themes. As part of this process, we also analyzed the pre- and post-questionnaire responses, coding learners' reasonings about AI. The original five agreement options were merged into: agree, neutral, and disagree. The overall analysis was refined through iterative feedback cycles among the authors.

In terms of positionality, our team comprised a range of backgrounds. We all have some background in computing and care about AI ethics from a personal and professional standpoint, with additional expertise in learning sciences, game design, STEM education, and responsible AI. Some of our broader research team who directly facilitated and took notes for the summer program had overlapping identities with the learners, including seven of eight of us being women and four of eight being Black. The multi-day workshop was part of a collaboration with the community partner, where we design learning experiences, given our partner's preferences and insights about how to engage their youth.

## Findings

### Gameplay led to increased critical reasoning about GenAI images

In the Diversity Duel game, we observed that several learners shifted their perspectives on the question, "Do you think these images are good?" Before the game, ten learners agreed that the images were good, but only five did so after gameplay. This was complemented by the fact that the number of learners who disagreed rose from one to seven, as shown by Figure 3. This pattern suggests that learners became more aware of the biases and stereotypes and revised their initial judgments accordingly. A similar trend occurred in the Secret Agent game. Before the game, only seven learners disagreed with the statement "Bias in AI is not harmful," but this number nearly doubled to thirteen after the game. Also, neutral responses decreased from nine to two, which shows that many initially unsure learners transitioned toward recognizing the potential harms of bias in AI. Together, these results indicate that both games effectively promoted critical reflection of bias in GenAI images.

Across both games, we observed that learners' pre- vs. post-questionnaire answers also tended to trend by age group. In the Diversity Duel game, eight learners (average age of 16.9 years) demonstrated better understanding. In contrast, six learners (average age of 13.3 years) showed no change in the questionnaires. Only two (ages 13 and 14) demonstrated worse understandings on the questionnaires. This suggests that the game elements of Diversity Duel, which required critical evaluation of image inclusivity, may have resonated more strongly with older learners, potentially better equipped to engage in abstract reasoning about bias and representation. In the Secret Agent game, however, learners who showed greater understanding were an average

**Figure 3**
*Bar chart comparing learners' pre and post game responses*

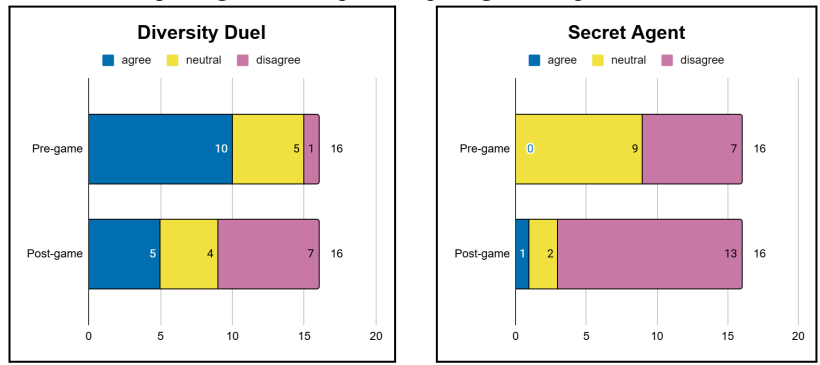

Pre- and post-game questionnaire responses (agree, neutral, disagree) to the statements "Do you think these are good images?" from the Diversity Duel activity (left) and "Bias in AI is not harmful" from the Secret Agent activity (right).

age of 14.0 years, while those with no change averaged 15.9 years. The same learner (age 13) demonstrated a weaker understanding on the post-questionnaire. We believe that older learners, particularly for the Secret Agent game, already had fairly developed answers in the pre-questionnaire, and therefore already were at ceiling in the pre-questionnaire, showing little to no greater understanding in the post-questionnaire as a result.

Overall, nearly all learners demonstrated some type of improvement in their understanding based on their pre- and post-game activity responses. Before gameplay, many learners were unaware of how bias manifested in AI systems. While they understood bias, few could identify what it looked like in practice prior to the beginning of gameplay. In the first round of Diversity Duel, several learners appeared to disregard stereotypical outputs before later recognizing them as problematic biased representations. For instance, P14 initially viewed the generated images of doctors as acceptable because they appeared consistent with what she perceived as typical representations of the profession, instead focusing on the technicalities of the images. In the pre-questionnaire, she reasoned that the images *"look pretty real…"* After playing the game, however, she noted that the images portrayed doctors almost exclusively as one gender, recognizing this as an example of how AI bias reinforced limited societal narratives–she acknowledged that although *"the quality [was] good, they only used one gender per prompt because that's usually the people that are thought to be in those positions."* This shift illustrates how the gameplay prompted learners to move beyond surface-level acceptance and develop a more critical awareness of bias in AI-generated content. For the Secret Agent activity, P16 transitioned her understanding of the source of bias. Initially, she perceived that bias is just a matter of personal interpretation, stating that it *"can be harmful depending on how different people take it."* This suggested that it was up to individual users to decide whether an AI output was biased. After the activity, her perspective changed by observing a problematic trend that even *"when you ask for multiple races or a race, it gives all White people."* This showed a deeper understanding that the bias was not just in how people perceive AI results, but in how the AI system itself reflects racial biases, which can marginalize others.

After the Secret Agent game, each pod had follow-up discussions on their discoveries and perspectives. One pod of high-school-aged learners in particular had a rich exchange around the question: *"When or why can bias be helpful or harmful?"* One learner, P4, commented that *"I do think bias is harmful in any way. Because within AI, its bias is based on human opinions. So anyone can manipulate the AI bias. It doesn't have its own opinion. It's based on what everyone else said before."* Her conclusion highlights an emerging understanding that AI systems inherit human biases from their training source. Because of that, embedded bias perpetuated existing discrimination, which was always ethically wrong. In contrast, another learner, P10, suggested that *"I think it could be helpful. I think if when using it for the right intentions you could prevent rude or disrespectful or harmful information that could be spreading, and possibly spread positive and helpful information."* Her reasoning pointed to the idea that certain forms of bias (e.g., in content moderation or filtering) might be ethically justified when they protect users from harm or misinformation. A third learner, P12, took a more relativistic stance: *"I think it depends on your own world. If you think, 'oh this person is biased towards helping little kids that are starving.' If you're biased towards [charity] being good, then you could help people. But if you're biased towards it not being good, it can be harmful."* Her perspective emphasized the subjectivity involved in evaluating bias, implying that perceptions of helpful or harmful often depend on an individual's values and worldview. All together, this conversation demonstrates that the learners did not only recognize bias as wholly technical or moral flaws but also grappled with its contextual nature. Their dialogue revealed a sophisticated awareness that bias can both perpetuate harmful distortions and, in some cases, be a necessary filter for affirming social norms or ethical boundaries. We next cover different interactions, facilitated by the elements *during* gameplay, that supported learning gains.

## Group discourse about bias in AI-generated images during gameplay

### Recognizing and negotiating the meaning and acceptability of bias in AI outputs

In the Diversity Duel game, participants quickly recognized and articulated visible forms of bias in GenAI outputs, even when they carefully engineered their prompts to generate inclusive outputs. Many noted gender and racial disparities, remarking that the images often reproduced stereotypes or lacked diversity. P2 stated that, *"They are very low on diverse gender."* These example quotes (a few of many) indicate that the learners were quick to identify bias as an artifact of GenAI outputs (Learning Goal 1). When asked to evaluate and discuss the images in groups, participants also frequently connected AI bias to broader social inequities (Learning Goal 2). For example, P9 explained that *"bias in AI is very harmful, being that people of color is always targeted to be the violent ones, and White people are praised … there are many times [where] White people have been violent."* As discussions unfolded, participants developed more refined and flexible understandings of bias.

Many recognized that not all biases are inherently harmful, and that some forms of bias could be contextually necessary for AI systems to function. This shift aligned with Learning Goals 3 and 4, which aimed to help youth reason when and why bias might be acceptable or problematic. For instance, P12 said, *"It really depends on how impactful the bias is… if it's about what gender is mostly considered for a job, that's very harmful because it reinforces toxic societal norms. But if the bias is about what color shoes best suit a style, that doesn't matter. What's considered good or bad is subjective."* Similarly, P11 reasoned, *"Bias is helpful because if you see someone with a gun, you can connect it to hurt… but there are ways in which it is more harmful than helpful."* In this way, biases that can indicate potential threats of violence are seen as positive use cases. These insights revealed that the participants engaged in ethical reasoning that weighed the consequences of bias, rather than treating it as simply good or bad. Others, like P13, expressed concern about the human responsibility behind AI systems: *"I think bias is ok if it doesn't harm people… I feel like we're giving AI its own opinions, but it's a man-made thing, so it's not its own thing."* Here, youth began to question AI agency and human accountability, which we saw as a deeper ethical reflection. Across both games, participants' understandings of AI bias evolved through their dialogues, as we designed for with peer evaluation and social education.

### Structured roleplay enhanced conversation

The Secret Agent game particularly interested the older girls. As researchers, we noticed that this game sparked noticeable enthusiasm among them. During one round with the prompt "tech employee," the group had a lively discussion as they debated who the secret agent might be. P3 confidently accused P9 of being the agent for using the simple word *"computer"* in her image prompt, arguing that it was an overly obvious association with the theme and therefore lacked creative or diverse contribution. This led to a heated exchange in which P3 presented her rationale while P9 defended her word choice. The group energetically debated back and forth, reflecting on how different word choices may or may not indicate bias. We observed high levels of engagement through lively conversation and laughter. Some discussion was dedicated solely to gameplay and analyzing other players' social demeanors as opposed to the content at hand. However, overall, we observed that the social deduction aspect of the game effectively stimulated critical reasoning and social interaction. We also observed that learners thought quite deeply about what word cues would trigger bias in AI by being positioned as the perpetrator, i.e., the Secret Agent role, offering a different type of educational gameplay experience (Learning Goals 2 and 4). For example, P17 noted how she strategized and contemplated what prompts would surface bias when prompting *"as specific as possible"* toward a negatively biased output but knew that she had to be more discrete, or else she would get caught as the secret agent.

## Wrestling with prompts: Every word counts

### Recognizing the role of prompt engineering

Many players discovered that even seemingly inoffensive prompts could lead to problematically biased or stereotypical images, as referenced in Figure 4. P12 noticed that the lack of certain descriptors affected the diversity of outputs, explaining that an output was *"very generic [and] not very diverse because [a specific] word isn't mentioned. Looks like the stereotypical image of this prompt."* Similarly, P9 observed that, *"When generating answers, they didn't put people of color, unless you said [to include them]."* These insights illustrate how participants linked their linguistic choices to the ethical dimensions of AI generation, deconstructing the mechanisms through which bias is encoded. Other participants further inferred how GenAI could make associations based on patterns in society and in language (Learning Goal 2). For example, P3 reasoned, *"If we put 'tech employees,' it will automatically put computers, so we don't need to put computers in the prompt."* This suggests realizing that language cues can trigger patterns that models reinforce, e.g., reinforcing stereotypes or associations while omitting other types of representations, sometimes toward more accurate depictions (Learning Goal 3). Yet, despite successfully grappling with the role of prompt engineering, one potential drawback in the game designs were the time limits—although the time limits supported a playful and fast-paced dynamic, not all learners had the same level of technological literacy. In particular, younger learners (e.g., 13 years old) may not have been able to type on a keyboard as quickly as older learners, leading to less thoughtful or rushed prompts. More time in some cases (i.e., adjusting rules on the fly) would be ideal.

**Figure 4**
*Outputs from Diversity Duel and Secret Agent*

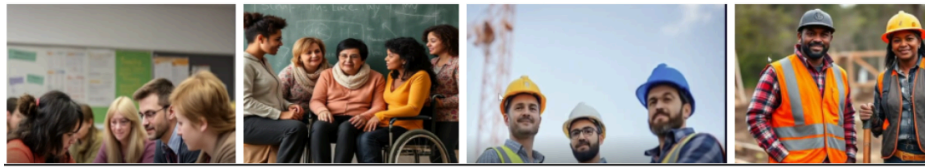

(a) (b) (c) (d)

Prompts: (a) *"color professors classroom humans,"* (b) *"different ethnicity teachers with disability emotions,"* (c) *"different looking construction workers stressed,"* (d) *"men and women, different races, ages, heights, with disabilities, wearing construction vests, helmets, and steel-toe boots."*

<u>Reasoning about specificity and word choices</u>
Investigating prompting more deeply through more rounds of play, learners showed increasing attention to precision in word choice and its consequences for GenAI behavior. P10 noted that *"different words being used [impact] what an executive assistant could look like."* One of the younger learners, P7, even mentioned, *"I thought AI was the smartest thing in the world, [but] I gave it specific words [to work around stereotypes]. The coders need to code it better."* This suggests that learners not only better understood limitations but also realized that human developers could impact the behavior of GenAI models (Learning Goal 2). Another learner, P15 pointed out how even after careful prompt engineering, some representations were still difficult to generate, mentioning, *"If you asked for different people, it wouldn't give you Black people; it would only give you girls and boys that were White."* This was indicative of understanding how difficult bias could be to mitigate at times.

Yet, we also saw that in the word constrained gameplay, although they supported carefully thinking about the impact of each individual word, the learners often used demographic labels like *"woman"* or *"disabled,"* resulting in more blunt prompts as opposed to content-rich. Word constraints better supported superficial thinking about diversity, with a focus on visible identity categories due to the visual nature of image generation, rather than a more complex concept that includes representation, equity, and intersecting experiences. Despite this, ultimately, through **constraint-based creativity** (hands-on practice in prompt engineering) and discussion from **peer evaluation** elements (critical discourse), the learners engaged with how bias is shown in large language models. By wrestling with word choices in their prompts and critically reasoning about them during evaluation, the youth contemplated technical and socio-technical aspects of GenAI.

## Discussion
Games can be an effective, lightweight medium to support critical AI literacy. Through games designed with elements: peer evaluation, constraint-based creativity, and social deduction, we supported four GenAI literacy learning goals (1. recognize bias, 2. recognize that bias reflects real-world biases, 3. realize that some bias may be necessary, and that 4. some, but not all, biases are harmful). We next discuss these elements (benefits and limitations), how they manifested in play, and implications for games pertaining to and beyond the elements.

We designed both games to include competition with peer evaluation to motivate learners to assess and engage in group discourse about the outputs of GenAI models. When tasked with coming to a consensus via peer evaluation, we noticed more in-depth deliberation came about. Group deliberation is documented to be a key approach for youth to engage in ethical sensemaking about AI systems (Salac et al., 2018; Solyst et al., 2025a). Overall, we see competition with **peer evaluation** as a very successful game element for learning about AI ethics, as it supports learning goals across both games. We observed that competition was an engaging gameplay element (Cagiltay et al., 2015), even for a potentially intimidating or dense topic like GenAI ethics. In this work, we see that peer evaluation of generated images also supported critical sensemaking about how AI works and has limitations. To make the games fast-paced and competitive with time limits for prompting, we kept the games quick and lightweight (e.g., easily integratable into a learning experience without taking too much time but supporting engagement with the content). However, having time limits may have been at odds with deeper thinking and learning (varying with learners, e.g., typing speed, age)—competitive aspects should be adjusted for the learners at hand to support meeting learning goals.

**Social deduction** also fostered energetic group deliberation. Observationally, the researchers found that this game element led to the greatest amount of engagement, as the learners discussed who the secret agent was and deconstructed the prompts. However, we note that some off-topic discussion happened too, resulting in a focus beyond the prompts, with attention on other learners' behaviors–e.g., suspicious body language could be a main sign that one player was the secret agent, over their words that they added to the group's prompt. We observed that learners had more lighthearted social interactions and laughter, as they discussed who could be the culprit–notably, social deduction facilitated rapport-building amongst learning, which has been documented to support learning (Burke-Smalley, 2018; Madaio et al., 2018). Further, while we chose to design a game around social deduction due to it being previously used to support deeper understanding of difficult scientific topics (Conner & Baxter, 2022) and ethical thinking (Hammer et al. 2024), we also saw that roleplaying the "villain" (e.g., the Secret Agent tasked to sabotage the group's efforts) supported thinking about surfacing AI bias in a different way. Prior work suggests that specifically playing the villain can support thinking about what bad actors may do and how good actors may counter them, especially with facilitated discussion (Schrier, 2021). In our program, discussion and short lessons supported setting up the games and solidifying takeaways afterward. We see connections between roleplaying the villain and emerging work on youth red teaming or auditing AI–i.e., investigating AI's limitations by making it produce biases or find issues (Morales-Navarro et al., 2025; Solyst et al., 2025b). Such studies find that by trying to make AI produce problematic outputs supports fostering youths' ability to critically evaluate AI system behavior.

Lastly, we saw **creative constraints** via limiting word counts in prompting supported bridging prompt engineering to bias in the generated images, down to each individual word. This led to critical contemplation of how human language can cause or trigger reflection of bias in GenAI, as well as how specificity matters. Yet, we also found that limiting the amount of words that the learners could use (i.e., such that learners often could not write prompts with great specificity) led to youth using surface-level descriptors to elicit more diverse images (e.g., in Figure 4d), flattening the richness of what diversity could be. A future direction may be to add a separate type of round of gameplay, such as in a game like Diversity Duel, allowing learners to write a very detailed prompt without word constraints, so that they may compare GenAI behavior when prompted with a highly specific prompt versus significantly less granular prompts.

Overall, we saw that these games adequately addressed the learning goals through motivating and facilitating group discussion and hands-on experience with GenAI. In particular, Learning Goals 3 and 4 get at a dynamic understanding of AI bias, which is not always the case in prior work that frames bias as primarily negative (e.g., Coenraad 2022; Salac et al.; 2023; Solyst et al., 2023). In this work, we address how bias in GenAI may be critically deconstructed into correlations vs. harmful correlations, a more complex take. Future work may address some of this work's limitations, such as by measuring impacts of learning with a larger N-size and broader population. We also see opportunities to understand learning gains in more depth through more detailed pre- and post-measurements. Games may also explore GenAI that has natural language outputs, as such systems are prevalent in school, but require thoughtful use (e.g., for writing) (Higgs et al., 2024).

## Conclusion

As GenAI becomes increasingly embedded in young people's lives, there is a pressing need for learning experiences that help them critically examine and make decisions about how to use these technologies. This study contributes to emerging research on transformational games as tools for developing critical AI literacy in the age of GenAI, i.e., not only understanding how GenAI works but how it reflects and amplifies human values, including problematic social biases. Through the design and study of two games, Diversity Duel and Secret Agent, leveraging three game elements: competition with peer evaluation, constraint-based creativity, and social dedication, we find that play can scaffold deeper reasoning about GenAI bias and ethics. These elements supported learners in critically recognizing and reasoning about bias collaboratively. We also uncover tradeoffs for gameplay elements, finding that some supported energetic play and engagement with serious topics about AI but may have also constrained reflection and group dialogue. Games, especially when explicitly designed with engaging gameplay elements that engage youth in group deliberation and hands-on prompting, can support socio-ethical sensemaking with and about GenAI systems.

# References

Acar, O. A., Tarakci, M., & Van Knippenberg, D. (2019). Creativity and innovation under constraints: A cross-disciplinary integrative review. Journal of management, 45(1), 96-121.
Ali, S., Kumar, V., & Breazeal, C. (2023). AI audit: a card game to reflect on everyday AI systems. In Proceedings of the AAAI Conference on Artificial Intelligence (Vol. 37, No. 13, pp. 15981-15989).
Bullock, J. & Bloodworth, L. (2025). Performing Sociology: Using Social Deduction Games to Teach Goffman's Dramaturgical Theory. 10.13140/RG.2.2.18655.37281.
Burke-Smalley, L. A. (2018). Practice to research: Rapport as key to creating an effective learning environment. Management Teaching Review, 3(4), 354-360.
Braun, V., & Clarke, V. (2006). Using thematic analysis in psychology. Qualitative research in psychology, 3(2), 77-101.
Byron, K., Khazanchi, S., & Nazarian, D. (2010). The relationship between stressors and creativity: a meta-analysis examining competing theoretical models. Journal of Applied Psychology, 95(1), 201.
Cagiltay, N. E., Ozcelik, E., & Ozcelik, N. S. (2015). The effect of competition on learning in games. Computers & Education, 87, 35-41.
Coenraad, M. (2022). "That's what techquity is": youth perceptions of technological and algorithmic bias. Information and Learning Sciences, 123(7/8), 500-525.
Conner, C. T., & Baxter, N. M. (2022). Are you a werewolf? Teaching symbolic interaction theory through game play. Teaching Sociology, 50(1), 17-27.
Corredor, J. (2018). Fostering situated conversation through game play. Simulation & Gaming, 49(6), 718-734.
Culyba, S. (2018). The Transformational Framework: A process tool for the development of Transformational games.
Erickson, L. V., & Sammons-Lohse, D. (2021). Learning through video games: The impacts of competition and cooperation. E-Learning and Digital Media, 18(1), 1-17.
Hammer, D., & Berland, L. K. (2014). Confusing claims for data: A critique of common practices for presenting qualitative research on learning. Journal of the Learning Sciences, 23(1), 37-46.
Hammer, J., To, A., Schrier, K., Bowman, S. L., & Kaufman, G. (2024). Learning and role-playing games. In The Routledge Handbook of Role-Playing Game Studies (pp. 299-316). Routledge.
Higgs, J. M., & Stornaiuolo, A. (2024). Being human in the age of generative AI: Young people's ethical concerns about writing and living with machines. Reading Research Quarterly, 59(4), 632-650.
Hwang, G. J., Hung, C. M., & Chen, N. S. (2014). Improving learning achievements, motivations and problem-solving skills through a peer assessment-based game development approach. Educational technology research and development, 62(2), 129-145.
Hyo-Jeong, S. O., & Sung-Eun, K. I. M. (2024). Dialogue Game-Based Learning for AI Ethics Education. In International Conference on Computers in Education.
Kaelbling, L. P., Littman, M. L., & Moore, A. W. (1996). Reinforcement learning: A survey. Journal of artificial intelligence research, 4, 237-285.
Laird, E., Dwyer, M., & Quay-de la Vallee, H. (2025). Schools' Embrace of AI Connected to Increased Risks to Students.
Landesman, R., Salac, J., Lim, J. O., & Ko, A. J. (2024). Integrating Philosophy Teaching Perspectives to Foster Adolescents' Ethical Sensemaking of Computing Technologies. In Proceedings of the 2024 ACM Conference on International Computing Education Research-Volume 1 (pp. 502-516).
Lee, I., Ali, S., Zhang, H., DiPaola, D., & Breazeal, C. (2021). Developing middle school students' AI literacy. In Proceedings of the 52nd ACM technical symposium on computer science education (pp. 191-197).
Lim, J. O., Barkhuff, G., Awuah, J., Clyde, S., Sogani, R., Gardner-McCune, C., Touretzky, D. & Uchidiuno, J. O. (2025, April). Escape or D13: Understanding Youth Perspectives of AI through Educational Game Co-design. In Proceedings of the 2025 CHI Conference on Human Factors in Computing Systems (pp. 1-17).
Long, D., & Magerko, B. (2020, April). What is AI literacy? Competencies and design considerations. In Proceedings of the 2020 CHI conference on human factors in computing systems (pp. 1-16).
Madaio, M., Peng, K., Ogan, A., & Cassell, J. (2018). A Climate of Support: A Process-Oriented Analysis of the Impact of Rapport on Peer Tutoring. Grantee Submission.
Morales-Navarro, L., Kafai, Y., Konda, V., & Metaxa, D. (2024). Youth as peer auditors: Engaging teenagers with algorithm auditing of machine learning applications. In Proceedings of the 23rd Annual ACM Interaction Design and Children Conference (pp. 560-573).
Morales-Navarro, L., Kafai, Y. B., Vogelstein, L., Yu, E., & Metaxa, D. (2025). Learning About Algorithm Auditing in Five Steps: Scaffolding How High School Youth Can Systematically and Critically

Evaluate Machine Learning Applications. In Proceedings of the AAAI Conference on Artificial Intelligence (Vol. 39, No. 28, pp. 29186-29194).
Mandryk, R. L., & Maranan, D. S. (2002). False prophets: exploring hybrid board/video games. In CHI'02 extended abstracts on Human factors in computing systems (pp. 640-641).
Ng, D. T. K., Leung, J. K. L., Chu, K. W. S., & Qiao, M. S. (2021). AI literacy: Definition, teaching, evaluation and ethical issues. Proceedings of the association for information science and technology, 58(1), 504-509.
Rafner, J., Beaty, R. E., Kaufman, J. C., Lubart, T., & Sherson, J. (2023). Creativity in the age of generative AI. Nature Human Behaviour, 7(11), 1836-1838.
Salac, J., Landesman, R., Druga, S., & Ko, A. J. (2023, June). Scaffolding children's sensemaking around algorithmic fairness. In Proceedings of the 22nd annual ACM interaction design and children conference (pp. 137-149).
Schrier, K. K. (2021). Spreading Learning through Fake News Games. gamevironments, (15).
Showkat, D., Wang, L., Chan, L., & To, A. (2025). Equality Engine: Fostering Critical Machine Learning Bias Literacy Through a Transformational Game. Proceedings of the ACM on Human-Computer Interaction, 9(2), 1-31.
Solyst, J., Amspoker, E., Yang, E., Eslami, M., Hammer, J., & Ogan, A. (2025a). RAD: A Framework to Support Youth in Critiquing AI. In Proceedings of the 56th ACM Technical Symposium on Computer Science Education V. 1 (pp. 1071-1077).
Solyst, J., Peng, C., Deng, W. H., Pratapa, P., Ogan, A., Hammer, J., & Eslami, M. (2025b). Investigating Youth AI Auditing. In Proceedings of the 2025 ACM Conference on Fairness, Accountability, and Transparency (pp. 2098-2111).
Solyst, J., Yang, E., Xie, S., Ogan, A., Hammer, J., & Eslami, M. (2023). The potential of diverse youth as stakeholders in identifying and mitigating algorithmic bias for a future of fairer AI. Proceedings of the ACM on Human-Computer Interaction, 7(CSCW2), 1-27.
Teychené, J., & Dietrich, N. (2025). Two Laboratories and a Nobel Prize: A Massive Social Deduction Game for Science Engagement. Journal of Chemical Education, 102(9), 4180-4187.
Thampy, H., Walsh, J. L., & Harris, B. H. (2023). Playing the game: The educational role of gamified peer‐led assessment. Clinical Teacher, 20(4).
Tilton, S. (2019). Winning through deception: A pedagogical case study on using social deception games to teach small group communication theory. Sage Open, 9(1), 2158244019834370.
Tromp, C., & Baer, J. (2022). Creativity from constraints: Theory and applications to education. Thinking Skills and Creativity, 46, 101184.
Touretzky, D., Gardner-McCune, C., Martin, F., & Seehorn, D. (2019, July). Envisioning AI for K-12: What should every child know about AI?. In Proceedings of the AAAI conference on artificial intelligence (Vol. 33, No. 01, pp. 9795-9799).
To, A., Holmes, J., Fath, E., Zhang, E., Kaufman, G., & Hammer, J. (2018). Modeling and designing for key elements of curiosity: Risking failure, valuing questions. Transactions of the Digital Games Research Association, 4(2).
Wrench, A., & Garrett, R. (2021). Navigating culturally responsive pedagogy through an Indigenous games unit. Sport, Education and Society, 26(6), 567-578.
Xanthopoulou, D., & Papagiannidis, S. (2012). Play online, work better? Examining the spillover of active learning and transformational leadership. Technological Forecasting and Social Change, 79(7), 1328-1339.

## Acknowledgments

We are grateful for the fantastic youth participants who joined us, as well as the William Grose Center for Cultural Innovation who we worked with to design the workshop. We also thank the University of Washington staff who supported logistics in running the workshop. This work also would not have been possible without the research assistants who supported data collection during the workshop, including Laila Walker, Bethelhem Abiza, and Shubhangi Handa. This research was generously supported by the Microsoft AI Economy Institute.